\documentclass[aps,prx,showpacs,amsmath,twocolumn,amssymb,superscriptaddress,letterpaper]{revtex4}
\usepackage{times}
\usepackage{amsfonts}
\usepackage{mathrsfs}
\usepackage{graphicx}
\usepackage{dcolumn}
\usepackage{bm}
\usepackage{color}

\usepackage[colorlinks,bookmarks=false,citecolor=blue,linkcolor=red,urlcolor=blue]{hyperref}
\usepackage{multirow}

\usepackage{color}
\usepackage{siunitx}
\usepackage{bibunits}

\def\be{\begin{equation}}       \def\ee{\end{equation}}
\def\bea{\begin{eqnarray}}      \def\eea{\end{eqnarray}}

\begin{document}

\begin{bibunit}

\title{Magnetism in Quasi-One-Dimensional A$_2$Cr$_3$As$_3$ (A=K,Rb) superconductors}
\author{Xianxin Wu }
\affiliation{ Institute of Physics, Chinese Academy of Sciences,
Beijing 100190, China}

\author{Congcong Le}  \affiliation{ Institute of Physics, Chinese Academy of Sciences,
Beijing 100190, China}

\author{Jing Yuan}  \affiliation{ Institute of Physics, Chinese Academy of Sciences,
Beijing 100190, China}

\author{Heng Fan}  \affiliation{ Institute of Physics, Chinese Academy of Sciences,
Beijing 100190, China} \affiliation{Collaborative Innovation Center of Quantum Matter, Beijing, China}

\author{Jiangping Hu}\email{jphu@iphy.ac.cn} \affiliation{ Institute of Physics, Chinese Academy of Sciences,
Beijing 100190, China}\affiliation{Department of Physics, Purdue University, West Lafayette, Indiana 47907, USA}
\affiliation{Collaborative Innovation Center of Quantum Matter, Beijing, China}

\date{\today}

\begin{abstract}
We predict that  the recently discovered quasi-one dimensional superconductors, A$_2$Cr$_3$As$_3$(A=K,Rb),  possess  strong frustrated magnetic fluctuations and are nearby a novel in-out co-planar magnetic ground state.   The frustrated magnetism is very sensitive to c-axis lattice constant and can thus be suppressed by increasing pressure.  Our results qualitatively explain  strong non-Fermi liquid behaviors observed in the normal state of  the superconductors
as the intertwining between the magnetism and superconductivity  can create a large quantum critical region  in quasi-one dimensional systems and also suggest that the materials share similar phase diagrams and superconducting mechanism
with other unconventional superconductors, such as cuprates and iron-based superconductors.

\end{abstract}


\maketitle
One of major challenges in condensed matter physics is to understand the role of electron-electron correlation in  unconventional superconductors.  The effect of electron-electron interaction becomes more important  as the  dimension of a system is lowered.  Indeed, many unconventional superconductors discovered in the past  are   quasi-two dimensional(Q2D)  electron systems.  The superconductivity in these  unconventional superconductors  appears in a vicinity to a magnetically ordered state.
Magnetic fluctuations which are caused by electron-electron interaction have  been widely considered to  be responsible for superconductivity and many non-Fermi liquid behaviors in normal states.

While there are many representatives of Q2D unconventional superconductors,  it has been difficult to find one  in  quasi-one dimensional(Q1D) systems  even if the effect of the electron-electron correlation is expected to be enhanced further. The Q1D superconductors discovered previously, including Bechgaard salts\cite{Jerome1980,Wilhelm2001} , Tl$_2$Mo$_6$Se$_6$\cite{Armici1980} and  Li$_{0.9}$Mo$_6$O$_{17}$\cite{Greenblatt1984,Denlinger1999,Xu2009,Mercure2012}, are not attributed to 3d-orbital electrons which can exhibt strong electron-electron interaction.

\begin{figure}[t]
\centerline{\includegraphics[height=8.5 cm]{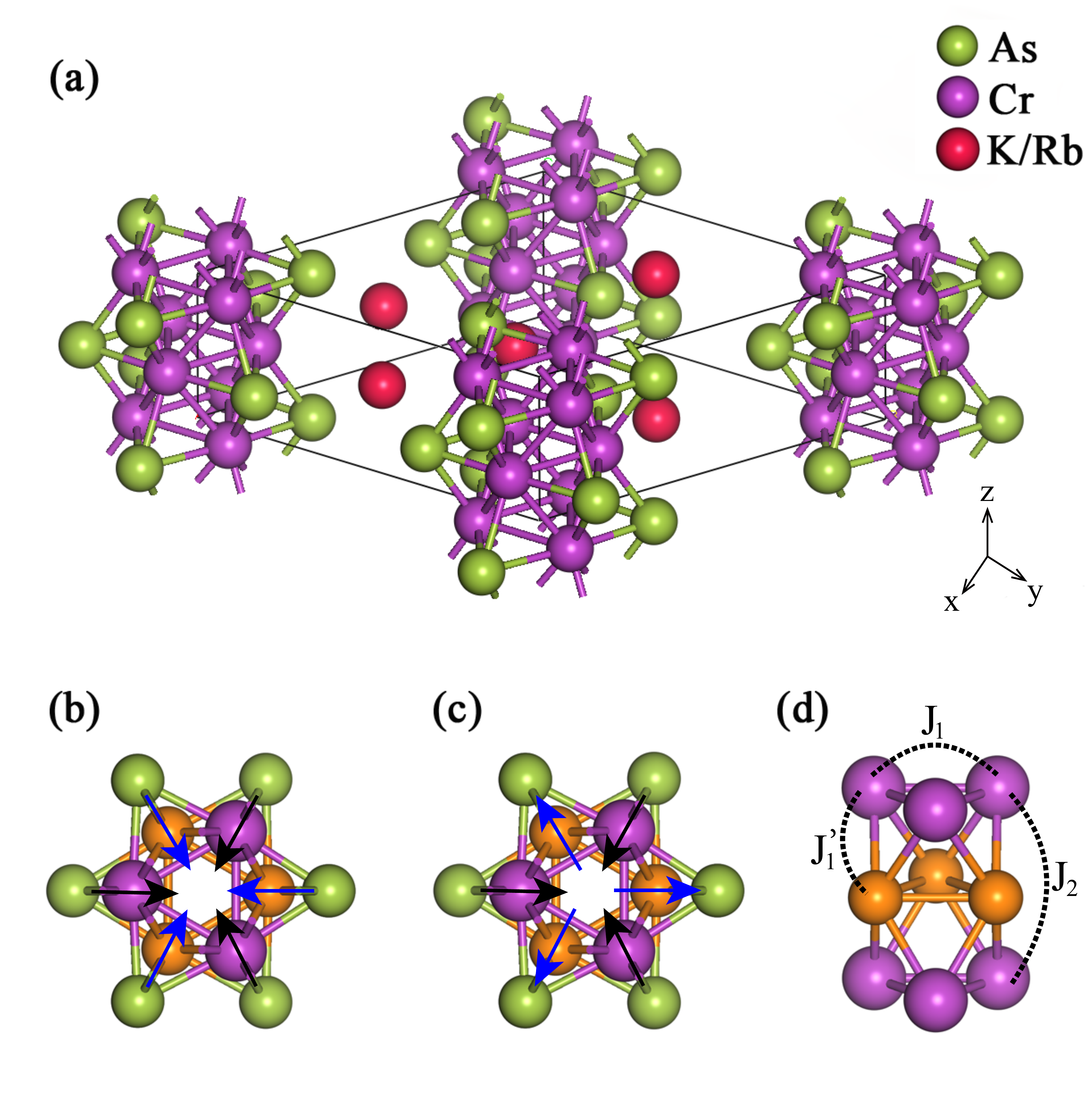}}
\caption{Schematic view of the structure of A$_2$Cr$_3$As$_2$ (a). (b) and (c) show the all-in and in-out noncollinear magnetic states. The exchange coupling parameters $J_1$, $J'_1$ and $J_2$ are defined in (d). The orange and purple spheres represent Cr1 and Cr2.
 \label{model} }
\end{figure}

Very recently, two novel Q1D materials K$_2$Cr$_3$As$_3$\cite{Bao2014} and Rb$_2$Cr$_3$As$_3$\cite{Tang2014} have been synthesized and found to be superconducting below the transition temperature 6.1 K and 4.8 K respectively. The structure of A$_2$Cr$_3$As$_3$ (A=K,Rb) is characterized by one-dimensional (Cr$_3$As$_3$) chains (Fig.\ref{model}(a)), which contain Cr$_6$ distorted octahedral clusters. The alkali metal ions are intercalated between the (Cr$_3$As$_3$) chains.  Both new materials show strong non-fermi liquid behaviors in normal states, as well as unconventional superconducting properties in superconducting (SC) states. Moreover, just like cuprates and iron-based superconductors, the electronic physics in these new materials are likely attributed to 3d-oribtals of Cr atoms. Therefore  the material may exhibit strong magnetism and electron-electron interaction.

In this paper, we show that the new materials can be a critical representative of Q1D unconventional superconductors where the superconductivity emerges in a vicinity to a novel magnetically ordered state.  We predict that the materials are characterized by   strong frustrated magnetic fluctuations and are nearby {\it a novel in-out co-planar(IOP) magnetic ground state}.  The magnetism can be  described by a minimum  effective model with three magnetic exchange parameters: the antiferromagnetic $J_1$ and $J'_1$ between two nearest-neighbor (NN) Cr  atoms and the ferromagnetic $J_2$ between two next NN Cr atoms along c-axis.  The frustrated magnetism is very sensitive to c-axis lattice constant and can thus be suppressed by increasing pressure.  The results suggest that the materials host a typical phase diagram similar to those of the Q2D unconventional superconductors, such as cuprates\cite{Lee2006} and iron-based superconductors\cite{Johnston2010}. The new materials can be  ideal systems to study the intimate  relations  between magnetism and superconductivity since  a Q1D model can be solved theoretically with high controllability.

It is known that  calculations based on the Density Functional Theory (DFT) can successfully identify the possible magnetic ground state in some correlated electron systems. Typical examples are the iron-based superconductors where  theoretical calculations consistently agree well with experimental measurements\cite{Mazin2008L,Cao2008,Ma2008,Dong2008,Yildirim2008L,Yin2008,Ma2009} even if the magnetic origin is still debatable\cite{Dai2012}.  Here, we deploy similar DFT calculations.  Our DFT calculations employ the projector augmented wave (PAW)
method encoded in Vienna \emph{ab initio} simulation package(VASP)
\cite{Kresse1993,Kresse1996,Kresse1996B}, and generalized-gradient approximation (GGA)\cite{Perdew1996} for the exchange correlation functional is used. Throughout this work,
the cutoff energy of 450 eV is taken for expanding the wave
functions into plane-wave basis. The number of these $k$ points are
$6\times6\times13$ for K$_2$Cr$_3$As$_3$ and Rb$_2$Cr$_3$As$_3$\cite{MonkhorstPack}.  The GGA plus on-site repulsion U method (GGA+U) in
the formulation of Dudarev {\it et al}.\cite{Dudarev1998} is employed to describe the
electron correlation effect associated with the Cr $3d$ states by
an effective parameter $U_{eff}$. The values of $U$=2.3 eV and
$J$=0.96 eV on Cr are adopted for our GGA+U calculations\cite{Mazin2007}.  We relax the lattice constants and internal atomic positions, where the plane wave cutoff energy is 600 eV. Forces
are minimized to less than 0.01 eV/\AA~ in the structural relaxation.

\begin{table}[bt]
\caption{\label{structure}%
Experimental and optimized structural parameters of K$_2$Cr$_3$As$_3$ using
GGA in the paramagnetic phase. Deviations between the optimized and experimental values are given in parentheses in \% .
}
\begin{ruledtabular}
\begin{tabular}{ccc}
& GGA & EXP \\
 \colrule
$a$(\AA)  & 10.113(+1.3) & 9.983 \\
$c$(\AA)  & 4.147(-1.98) & 4.230 \\
Cr1-Cr1/Cr2-Cr2(\AA)  & 2.498; 2.588 & 2.615; 2.691  \\
Cr1-As1/Cr2-As2(\AA) & 2.513; 2.494 & 2.51; 2.49 \\
Cr1-As2/Cr2-As1(\AA) & 2.522; 2.506 & 2.516; 2.506 \\
$\alpha_1$/$\alpha_2$(${}^\circ$) & 59.6; 62.51 & 62.8;65.4 \\
$\beta_1$/$\beta_2$(${}^\circ$) & 60.8; 60.87 & 62.8;62.9 \\
$\gamma_1$/$\gamma_2$(${}^\circ$) & 110.6; 111.6 & 114.4;115.2 \\
\end{tabular}
\end{ruledtabular}
\end{table}

Due to the asymmetric distribution of alkali metal ions and the absence of inversion symmetry, there are two kinds of Cr ions in K$_2$Cr$_3$As$_3$(Rb$_2$Cr$_3$As$_3$): Cr1 and Cr2. The Cr1 ions are surrounded by six inplane A ions and the inplane bondlength is 2.615(2.57)\AA.~ The Cr2 ions are surrounded by three inplane alkali ions and the inplane bondlength is a bit longer, 2.69(3.16)\AA. From the number of A ions surrounding Cr ions, it is expected that the Cr1As layer will obtain more electrons than the Cr2As layer.

The optimized and experimental structural parameters of K$_2$Cr$_3$As$_3$ are summarized in Table
\ref{structure}. We find that the lattice constants are comparable with experimental values but the Cr-Cr bond lengths are underestimated by 0.1 \AA.~ Thus, the Cr1-As1-Cr1($\alpha_1$)/Cr2-As2-Cr2($\alpha_2$), Cr1-As1-Cr2($\beta_1$)/Cr2-As2-Cr1($\beta_2$) and Cr1-As2-Cr1($\gamma_1$)/Cr2-As1-Cr2($\gamma_2$) angles are also underestimated. Similar cases have been also noted in the studies of iron based superconductors\cite{Yin2008,Singh2008,Mazin2008} and it may be related to the strong spin fluctuation, which is beyond DFT calculation. We adopt the experimental parameters in the following calculation unless otherwise specified.

The band structure, the density of states(DOS) and Fermi surfaces for K$_2$Cr$_3$As$_3$ have been calculated in Ref.\onlinecite{Jiang2014}.  Our results as shown in Appendix A are similar to their calculations.  In general,  the 3$d$ states of Cr are located from -2.5 eV to 2.0 eV and the states near the Fermi level are mainly attributed to Cr $d_{z^2}$, $d_{xy}$ and $d_{x^2-y^2}$ orbitals. The As $4p$ states mainly lie 1.0 eV below the Fermi level and
hybridize strongly with the Cr $3d$ states. Along $k_z$ direction, the bands are very dispersive due to the strong coupling between $d$ oribials along $z$ direction. The bands attributed to Cr2 $d_{xy}$ and $d_{x^2-y^2}$ orbitals are very close to half-filling. As the separation between chains is large, the inplane band dispersion is very small but not negligible. However, we want to point out an important feature that is ignored in Ref.\onlinecite{Jiang2014}.
Cr1 ions have more $d$ electrons than Cr2 ions, which is consistent with analysis from the crystal structure.  We will show later that this difference also results in different magnetic moments at Cr1 and Cr2 sites in magnetic states.

The band structure of Rb$_2$Cr$_3$As$_3$, shown also in Appendix A,  shares many common features with that of K$_2$Cr$_3$As$_3$, but they also differ in some details. As the radius of Rb is bigger than K, Cr1As layer can get more electrons and $d_{xy}$ and $d_{x^2-y^2}$ bands of Cr1 are fully occupied. Meanwhile, the $d_{xy}$ and $d_{x^2-y^2}$ bands of Cr2 are much less occupied. It leads to great difference between the 3D Fermi surfaces of the two materials. Another difference is that there is an additional electron Fermi surface near $A(0,0,\pi)$ point in Rb$_2$Cr$_3$As$_3$, which is attributed to $d_{xy}$ and $d_{x^2-y^2}$ orbitals of Cr2. Finally the $d_{xz}$ and $d_{yz}$ orbitals are much close to the Fermi level in Rb$_2$Cr$_3$As$_3$. The calculated N($E_F$) in K$_2$Cr$_3$As$_3$(Rb$_2$Cr$_3$As$_3$)  is 8.76(9.13) eV$^{-1}$/f.u.. The calculated Pauli susceptibility and specific heat coefficient are $\chi_0=$ 2.83$(2.95)\times 10^{-4}$ emu/mol and $\gamma=$ 20.7(21.5) mJ/($K^2$ mol). The calculated $\gamma$ is about only one third of the experimental value in K$_2$Cr$_3$As$_3$, suggesting strong correlation in these systems.

Now we focus on the magnetic properties. We consider four possible collinear magnetic states, the paramagnetic state (PM), the ferromagnetic (FM) state, the interlayer antiferromagnetic (AFM) state and the up-up-down-down ($\uparrow\uparrow\downarrow\downarrow$) magnetic state. Due to strong magnetic frustration, we also consider two additional co-planar antiferromagnetic states: all-in magnetic state  (Fig.\ref{model}(b)) and in-out magnetic state (Fig.\ref{model}(c)). We perform calculations with spin orbital coupling (SOC) and the calculated magnetic moments and the total relative energies of above magnetic states are summarized in Table.\ref{magnetic}. The IOP  state is the ground state in K$_2$Cr$_2$As$_2$(Rb$_2$Cr$_2$As$_2$), with a large energy gain of 39(762) meV/cell relative to the PM state. The magnetic moments are 0.90(1.75) and 0.94(2.34) $\mu_B$ on Cr1 and Cr2 sites, respectively. The initial FM state converges to a PM or AFM state, indicating that the interlayer magnetic coupling $J'_1$ as indicated in Fig.\ref{model}(d) is antiferromagnetic and relatively strong. In  the AFM state, the energy gain is 25(486) meV/cell and the three magnetic moments at Cr1 or Cr2 site are different. It manifests that the intralayer magnetic couplings among three Cr atoms $J_1$ is AFM. Thus,  strong magnetic frustration exists in A$_2$Cr$_3$As$_3$. Our results are different from those in Ref.\cite{Jiang2014} which missed to identify the IOP as the true magnetic ground state in their calculations  To confirm our results, we have performed all-electron calculations and find that relative energies and magnetic moments for collinear states are rather close to those in Table \ref{magnetic}.

The IOP state gains energy rapidly and becomes very robust if we  include  additional $U$, the onsite electron-electon correlation.  The results from GGA+U calculations  are given in Table.\ref{magneticU}.   The IOP  state in K$_2$Cr$_2$As$_2$(Rb$_2$Cr$_2$As$_2$)  has a large energy gain of 392(466) meV/cell relative to the AFM state. The magnetic moments are greatly enhanced to 2.49(2.76) and 2.56(2.92) $\mu_B$ on Cr1 and Cr2 sites as well. It is also important to note that   the magnetism is always found to be stronger in Rb$_2$Cr$_3$As$_3$  than in K$_2$Cr$_3$As$_3$. In the presence of $U$, as the calculation can be converged for different magnetic states, we can extract the magnetic exchange parameters from their energy differences\cite{Wu2012}. In the GGA+U calculations, the local magnetic moments on Cr are very close to each other. Thus, we can
can extract the magnetic exchange parameters from their energy differences within the Heisenberg Model. The energies contributed by
magnetic interactions in these five magnetic states per cell are written as,
\begin{eqnarray}
E_{FM}/S^2&=&6J_1+12J'_1+6J_2,\nonumber\\
E_{AFM}/S^2&=&6J_1-12J'_1+6J_2,\nonumber\\
E_{\uparrow\uparrow\downarrow\downarrow}/S^2&=&6J_1-6J_2,\nonumber\\
E_{all-in}/S^2&=&-3J_1+6J'_1+6J_2,\nonumber\\
E_{in-out}/S^2&=&-3J_1-6J'_1+6J_2.
\label{exchange }
\end{eqnarray}
Using the relative energies and the above equations, we can obtain four equations. But there are only three exchange parameters. For this overdetermined system of equations, we can obtain these parameters by using a least-squares technique. The estimated exchange couplings are $J_1=0.090$ eV/S$^2$, $J'_1=0.060$ eV/S$^2$, $J_2=-0.010$ eV/S$^2$ for K$_2$Cr$_3$As$_3$ and $J_1=0.075$ eV/$S^2$, $J'_1=0.045$ eV/$S^2$, $J_2=-0.020$ eV/$S^2$ for Rb$_2$Cr$_3$As$_3$ in the GGA+U calculations.   The NN exchange couplings   $J_1$ and $J'_1$ are strongly AFM. However, the next NN exchange coupling $J_2$ is ferromagnetic. The IOP state saves energy from all these effective exchange couplings.

\begin{figure}[t]
\centerline{\includegraphics[height=6 cm]{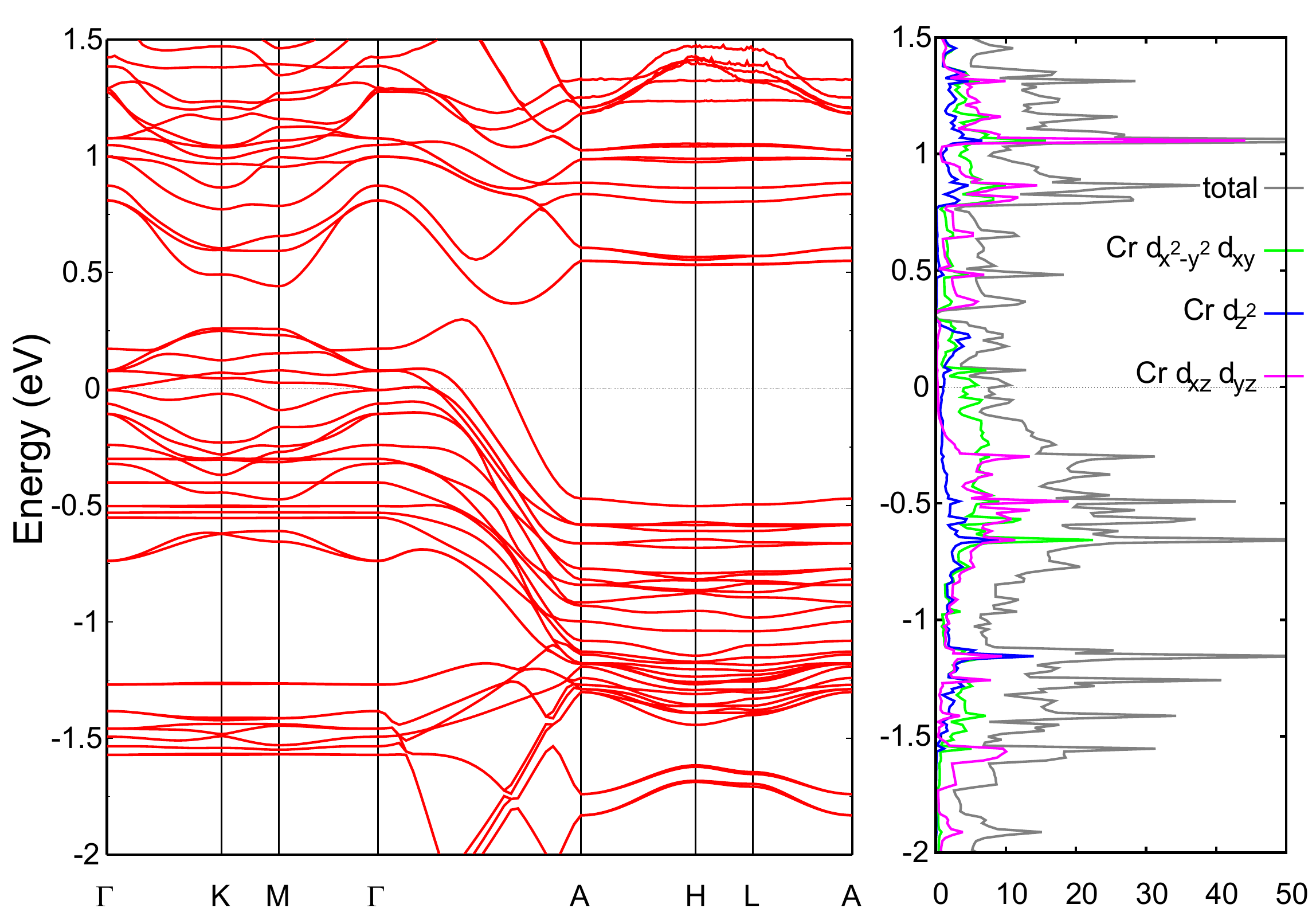}} \caption{Band structure and DOS for K$_2$Cr$_3$As$_3$ in in-out co-planar (IOP) magnetic state.
 \label{NCL} }
\end{figure}

 Fig.\ref{NCL}(a) and (b) show the calculated band structure and DOS in the IOP magnetic state of K$_2$Cr$_3$As$_3$. The material remains metallic in the presence of the static magnetic order.
The bands near the Fermi level split due to SOC and magnetic order but the orbital characters have little changes compared with those in PM state. This splitting depends on orbital characters of the bands: it is larger for $d_{xy}$,$d_{x^2-y^2}$ and $d_{xz}$, $d_{yz}$ bands than  for $d_{z^2}$ bands.

\begin{table}[bt]
\caption{\label{magnetic}%
The total energies for different magnetic states of A$_2$Cr$_3$As$_3$. They are given
relative to the total energy of the PM state and the unit is
meV/cell. The magnetic moments are given in $\mu_B$.}

\begin{ruledtabular}
\begin{tabular}{ccccc}
A$_2$Cr$_3$As$_3$& & relative energy (meV/cell) & $M_{Cr1}$  &  $M_{Cr2}$ \\
 \colrule
  \multirow{5}{*}{A=K}
&PM & 0  & 0  & 0  \\
&FM & to PM  &  & \\
&AFM & -25  & 0.43 &  0.53 \\
&$\uparrow\uparrow\downarrow\downarrow$ & to PM  & &   \\
&all-in& -3(to in-out)  &0.06   & 0.35\\
&in-out & -39  & 0.90  & 0.94  \\
 \colrule
 \multirow{5}{*}{A=Rb}
&PM & 0  & 0  & 0  \\
&FM & -589(to AFM)  & 0.40 & 2.29 \\
&AFM & -486  & 1.26 &  2.35 \\
&$\uparrow\uparrow\downarrow\downarrow$ & -525  & 0.01 & 1.83   \\
&all-in& -594(to in-out)  & 0.34  & 2.15 \\
&in-out & -762  & 1.75  & 2.34  \\
\end{tabular}
\end{ruledtabular}
\end{table}

\begin{table}[bt]
\caption{\label{magneticU}%
The total energies for different magnetic states of A$_2$Cr$_3$As$_3$ with GGA+U calculations. They are given
relative to the total energy of the AFM state and the unit is
meV/cell. The magnetic moments are given in $\mu_B$.}
\begin{ruledtabular}
 \renewcommand{\multirowsetup}{\centering}
\begin{tabular}{cccccc}
A$_2$Cr$_3$As$_3$& & relative energy (meV/cell) & $M_{Cr1}$  &  $M_{Cr2}$ \\
 \colrule
 \multirow{5}{*}{A=K}
&FM & +1495 &2.47  & 2.58 \\
&AFM & 0 & 2.48 &2.41 \\
&$\uparrow\uparrow\downarrow\downarrow$ & +854  & 1.50 & 2.44  \\
&all-in&+311 & 2.35   & 2.43 \\
&in-out & -392  & 2.49  & 2.56  \\
 \colrule
 \multirow{5}{*}{A=Rb}
&FM & +1045  & 2.09  & 2.84 \\
&AFM & 0 & 2.70 & 2.97 \\
&$\uparrow\uparrow\downarrow\downarrow$ & +792  & 2.72 & 2.90   \\
&all-in&+184 & 2.90   & 2.96 \\
&in-out & -466  & 2.76  & 2.92  \\

\end{tabular}
\end{ruledtabular}
\end{table}

\begin{figure}[t]
\centerline{\includegraphics[height=5.2 cm]{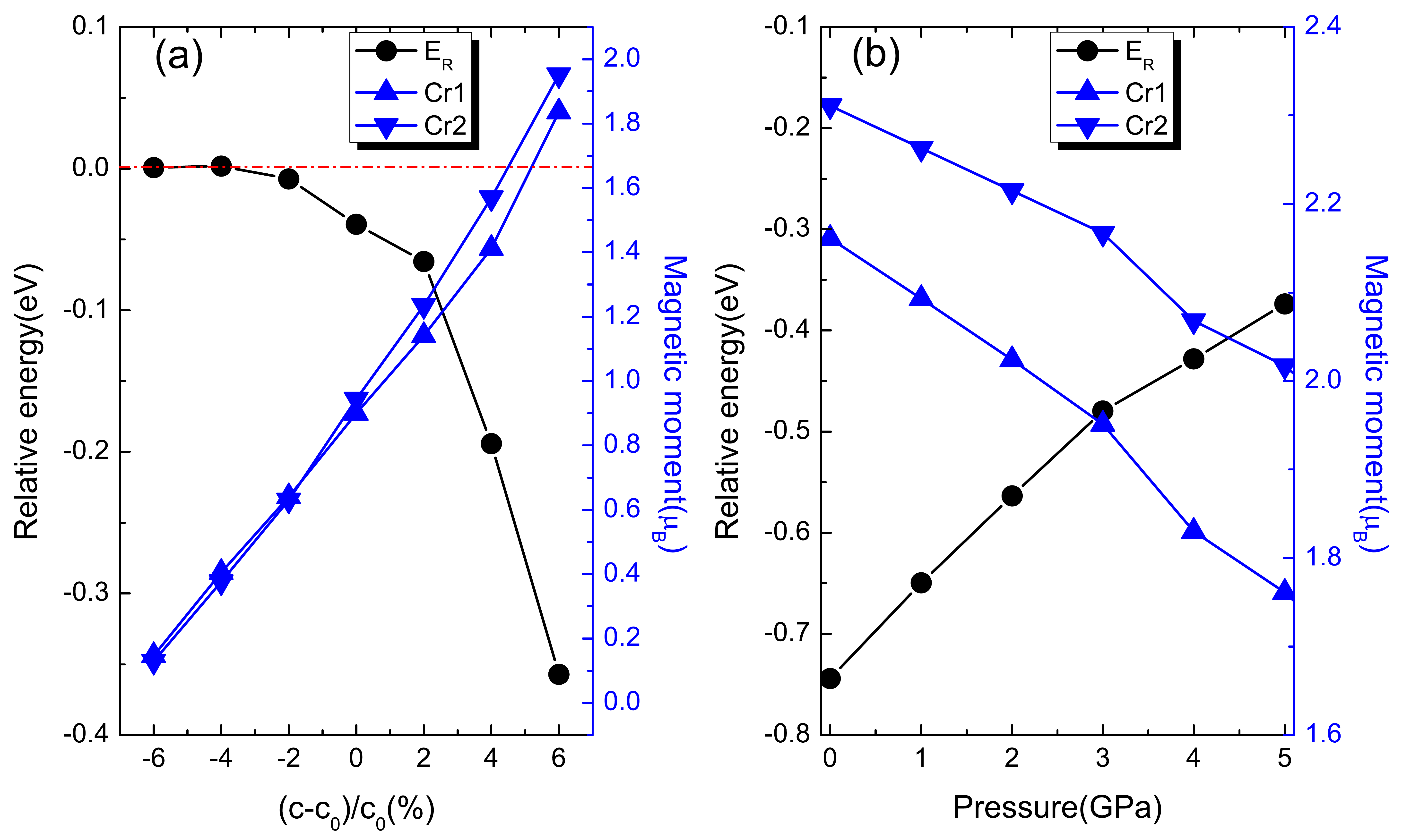}} \caption{The relative energies and magnetic moments as function of lattice constant $c$ (a) and pressure (b). In the case of pressure, we performed structure relaxation and GGA+U calculations.
 \label{pressure} }
\end{figure}

\begin{figure}[t]
\centerline{\includegraphics[height=3.8 cm]{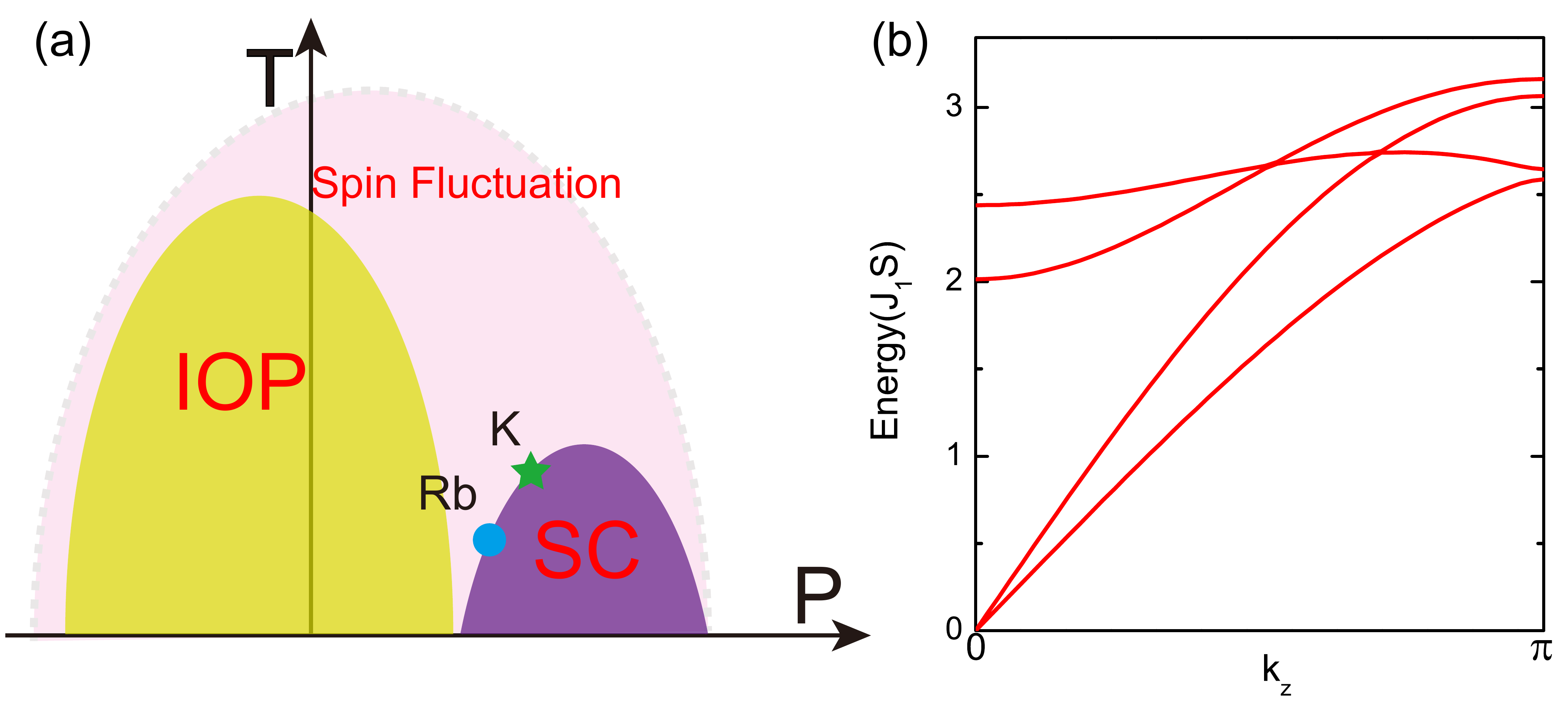}} \caption{(a) The conjectured  phase diagram with pressure of A$_2$Cr$_3$As$_3$. (b) The calculated spin-wave dispersion of A$_2$Cr$_3$As$_3$ along $k_z$ direction, with $J'_1$=0.7$J_1$ and $J_2$=-0.15$J_1$.
 \label{phasespinwave} }
\end{figure}

As the size of Rb atoms is much larger than the size of K atoms, the  stronger magnetism in Rb$_2$Cr$_3$As$_3$ than in K$_2$Cr$_3$As$_3$ suggests that applying pressure can suppress magnetic fluctuations.
To simulate the effect of pressure, we investigate the magnetism as a function of  the lattice constant $c$. Fig.\ref{pressure}(a) shows the calculated magnetic moments and relative energies of the IOP magnetic state(relative to PM state) as a function of $c$. As the decrease of $c$, which is equivalent to  the increase of pressure, the magnetic moments decrease, as well as the energy gains. When $c=4.06$ \AA,  the energy gain reaches  zero, indicating the vanish of IOP magnetic orders.  We also directly calculate the pressure effect using the  GGA+U calculation. Consistent results are obtained as shown in Fig.\ref{pressure}(b). These results allow us to conjecture a phase diagram of these new superconductors similar to the one shared by many Q2D unconventional superconductors as sketched in Fig.\ref{phasespinwave}(a). The  non-Fermi liquid behaviors observed in experiments can be naturally explained as a result of strong magnetic fluctuations in the critical region.

We can construct a minimum effective Hamiltonian to describe the magnetism for a Q1D chain in A$_2$Cr$_3$As$_3$ with the above mentioned magnetic exchange couplings.  The Hamiltonian is
 \begin{eqnarray}\label{hm}
 H_{M}&=&J_1\sum_{i,\alpha\beta}\textbf{S}_{i\alpha}\cdot \textbf{S}_{i\beta}+J'_1\sum_{\langle ij\rangle,\alpha\beta}\textbf{S}_{i\alpha}\cdot \textbf{S}_{j\beta}\nonumber \\
 &+&J_2\sum_{\langle\langle ij \rangle\rangle,\alpha\beta}\textbf{S}_{i\alpha}\cdot \textbf{S}_{j\beta},
 \end{eqnarray}
 where $\alpha,\beta$ label sublattices and $\langle ij\rangle$ and $\langle\langle ij \rangle\rangle$ denote the NN and next NN pairs. With $J_1, J_1'$ being AFM and $J_2$ being FM,   $H_M$ has  the  IOP magnetic ground state as a classical spin model or in the large S-limit.   The magnetic excitations can be calculated by   employing the Holstein-Primakoff transformation within the linear spin-wave approximation. As the 1D chain described by $H_M$ has a $C_{3v}$ symmetry with six Cr atoms  in the unit cell, there must be three acoustic  and three optical spin excitations. Each of these three modes must be composed of one $A_1$ and two degenerated $E$ modes.  The detailed analytic results of the spin waves are given in Appendix \ref{S3}.   A typical spin wave spectrum is plotted in Fig.\ref{phasespinwave}(b). The acoustic spin wave dispersion shows a linear dependence with small $k_z$, which indicates the AFM nature of the magnetic exchange couplings.

Although there is no direct experimental measurement on magnetic properties in these materials,  the magnetic exchange coupling parameters obtained from our calculation
are consistent with recent experimental measurements in CrAs\cite{Shen2014}, a MnP-type orthorhombic crystal structure and also a superconductor under pressure\cite{Wu2014}.  The magnetic order in CrAs is  a double helimagnetic structure. Along the one-dimensional helimagnetic structure,  the measurements  have shown that it is  AFM between two NN Cr atoms that has  a Cr-As-Cr angle $\sim 72^\circ$  and  FM between two next NN Cr atoms that has a Cr-As-Cr angle $\sim 123^\circ$. These two magnetic exchange couplings in CrAs resemble $J_1( J'_1)$ and $J_2$.  It is interesting to note that the exchange couplings between two Cr atoms are against the simple Goodenough-Kanamori-Anderson (GKA) rules which would suggest opposite signs for $J_1 (J'_1)$ and $J_2$. Such a violation suggests that multi-magnetic mechanisms may work together  in CrAs based structures and   $J_2$ may include  significant contribution from the double-exchange mechanism since the average valence of Cr atoms in A$_2$Cr$_3$As$_3$ is $2.3$.

We have ignored  the Dzyaloshinskii-Moriya(DM) term that can be induced by spin orbital couplings in the absence inversion center between two Cr atoms. In CrAs, the DM term is important in determining the incommensurate wavevector in the double helimagnetic structure. However, in  A$_2$Cr$_3$As$_3$, the effect of the DM is greatly reduced as  the Cr atoms form a tight octahedral cluster.  This is consistent with   our calculation results that  magnetic moments tend to be in the $xy$ plane.

Our results suggest that similar to cuprates and iron-based superconductors, the magnetic fluctuations are  responsible for the superconductivity in  A$_2$Cr$_3$As$_3$. If this is the case, it is interesting to ask what type of pairing is favored from the specific magnetic fluctuations predicted here.   We argue that the favorite pairing is likely to be spin-triplet pairing from the magnetic fluctuations. As shown in Appendix \ref{S1}, the Cr1 and Cr2 atoms make major contributions to different bands near the Fermi level. Thus, as the pairing is expected to be dominated by the intra-band pairing, in the real space picture the majority pairing takes place within each sublattice that only contains one type of Cr atoms.  The effective magnetic fluctuations  within each sublattice  along the the chain direction connected by $J_2$  are clearly FM from the above results. Therefore, if we ignore the parity breaking caused by alkaline atoms,  the pairing from the magnetic fluctuations is very likely to be a $p$-wave spin triplet as the correlation effect forbids  any onsite pairing. If we consider that the pairing is determined by the local FM exchange coupling\cite{Hu2012},   the induced gap function , $\Delta \propto sink_z$, which  should be  characterized by line nodes on the Fermi surface in the $k_z=0$ plane in reciprocal space.

In summary, we predict the Q1D   superconductors, A$_2$Cr$_3$As$_3$, are close to a novel in-out co-planar  magnetic ordered state and share a typical phase diagram similar to those of Q2D high temperature superconductors, cuprates and iron-based superconductors.  The prediction  qualitatively explains the non Fermi-liquid behaviors  observed in these systems and suggests  that the superconductivity in these systems are driven by electron-electron correlation effects.  We also predict that  $T_c$  can  be maximized in these systems by applying certain external pressure. The new materials can be an ideal Q1D system to understand the intimate relation between magnetism and superconductivity.

{\it Acknowledgments} We thank S. M. Nie and Q. Xie for the help on calculations. The work is supported by "973" program (Grant No.
 2010CB922904, No. 2012CV821400 and No. 2015CB921300),  the National Science Foundation of China (Grant No. NSFC-1190024, 11175248 and 11104339) and the Strategic Priority Research Program of  CAS (Grant No. XDB07000000). We also want to note that the authors in\cite{Jiang2014} have  checked our results
and have acknowledged to us that they missed the IOP ground state.

\clearpage
\pagebreak

\appendix
\section{Electronic structures of A$_2$Cr$_3$As$_3$} \label{S1}

\begin{figure}[b]
\centerline{\includegraphics[height=5 cm]{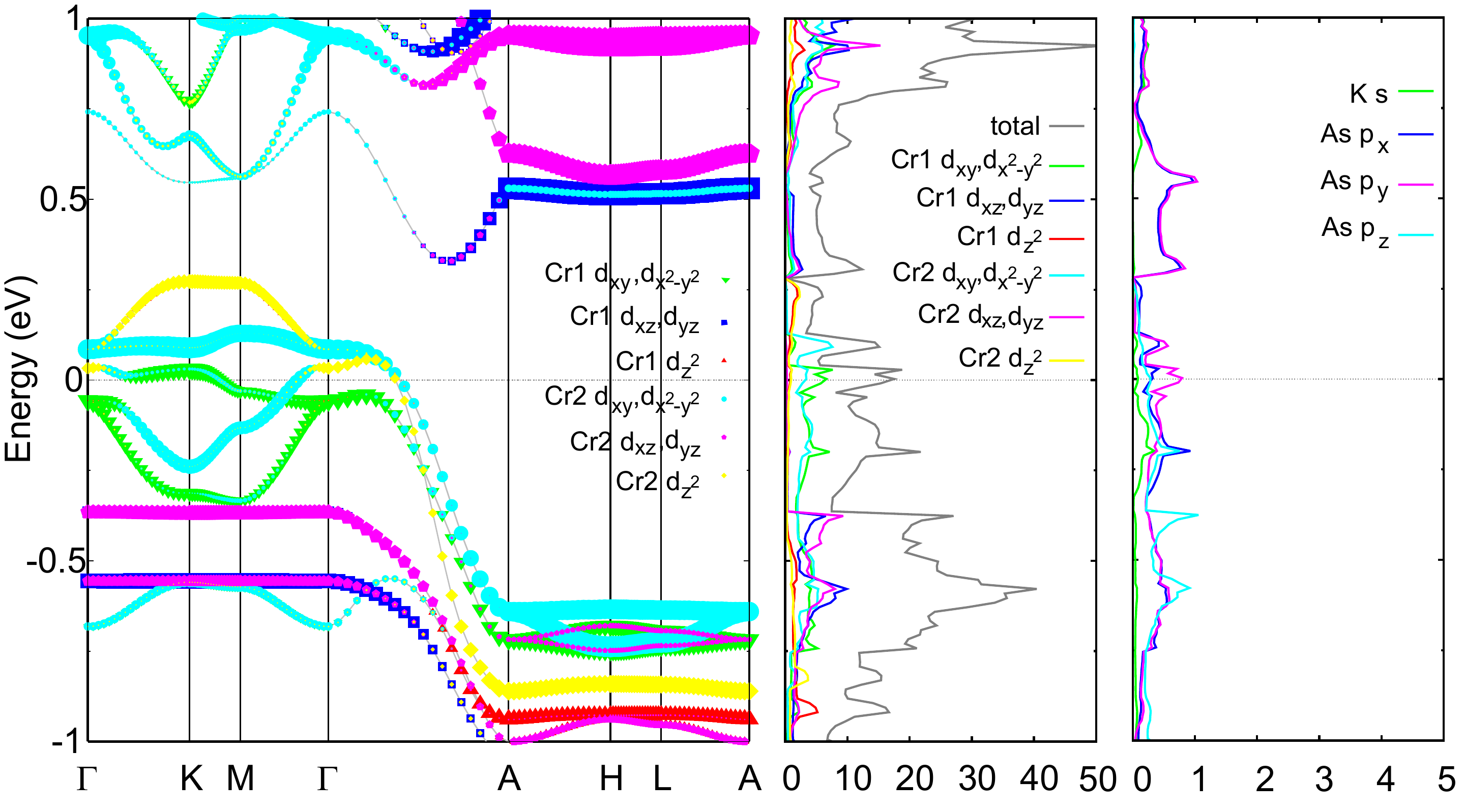}}
\caption{ Band structure and DOS of K$_2$Cr$_3$As$_3$ with experimental parameters in the paramagnetic state.
 \label{band} }
\end{figure}

\begin{figure}[b]
\centerline{\includegraphics[height=5 cm]{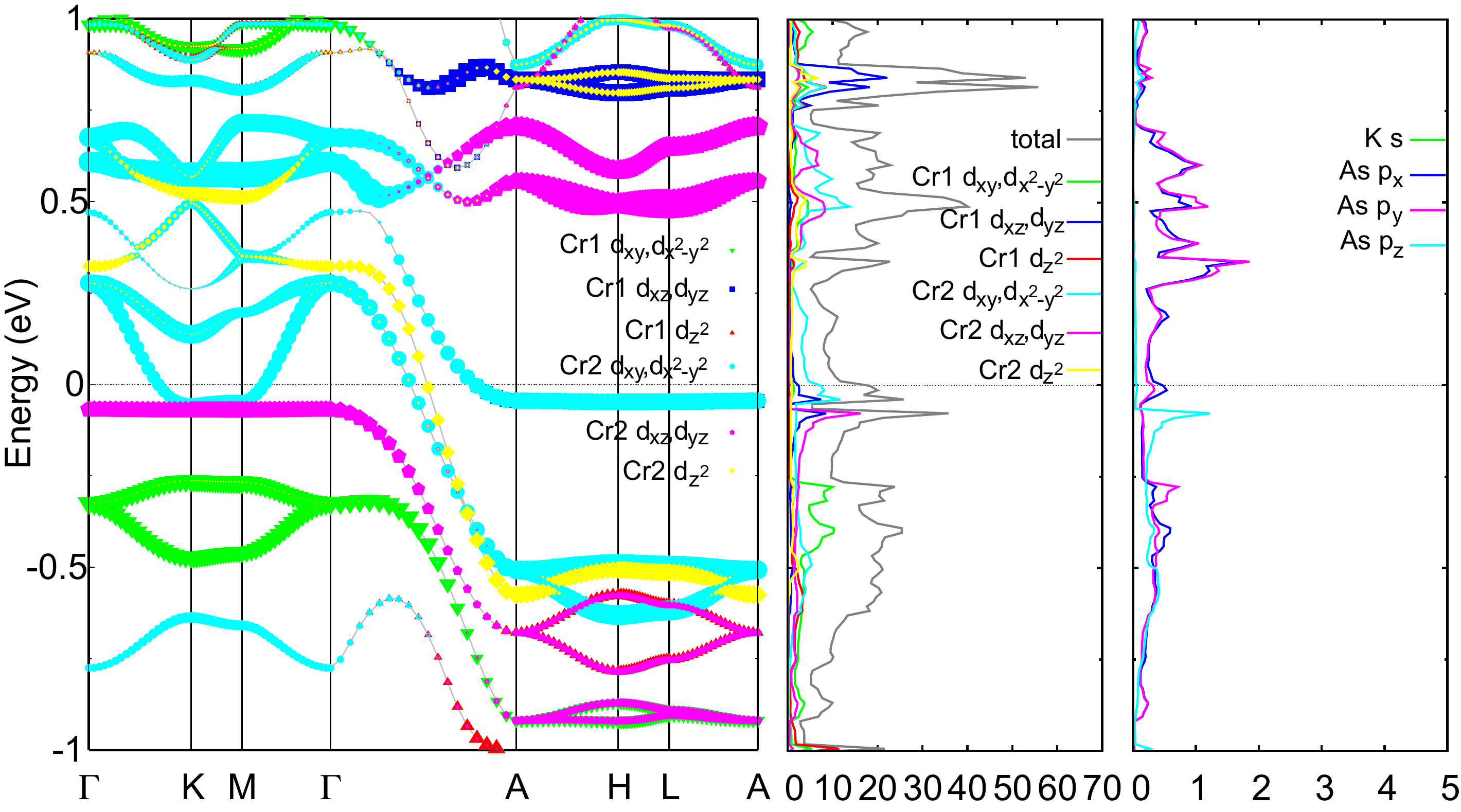}}
\caption{ Band structure and DOS of Rb$_2$Cr$_3$As$_3$ with experimental parameters in the paramagnetic state.
 \label{bandRb} }
\end{figure}

\begin{figure}[tb]
\centerline{\includegraphics[height=5cm]{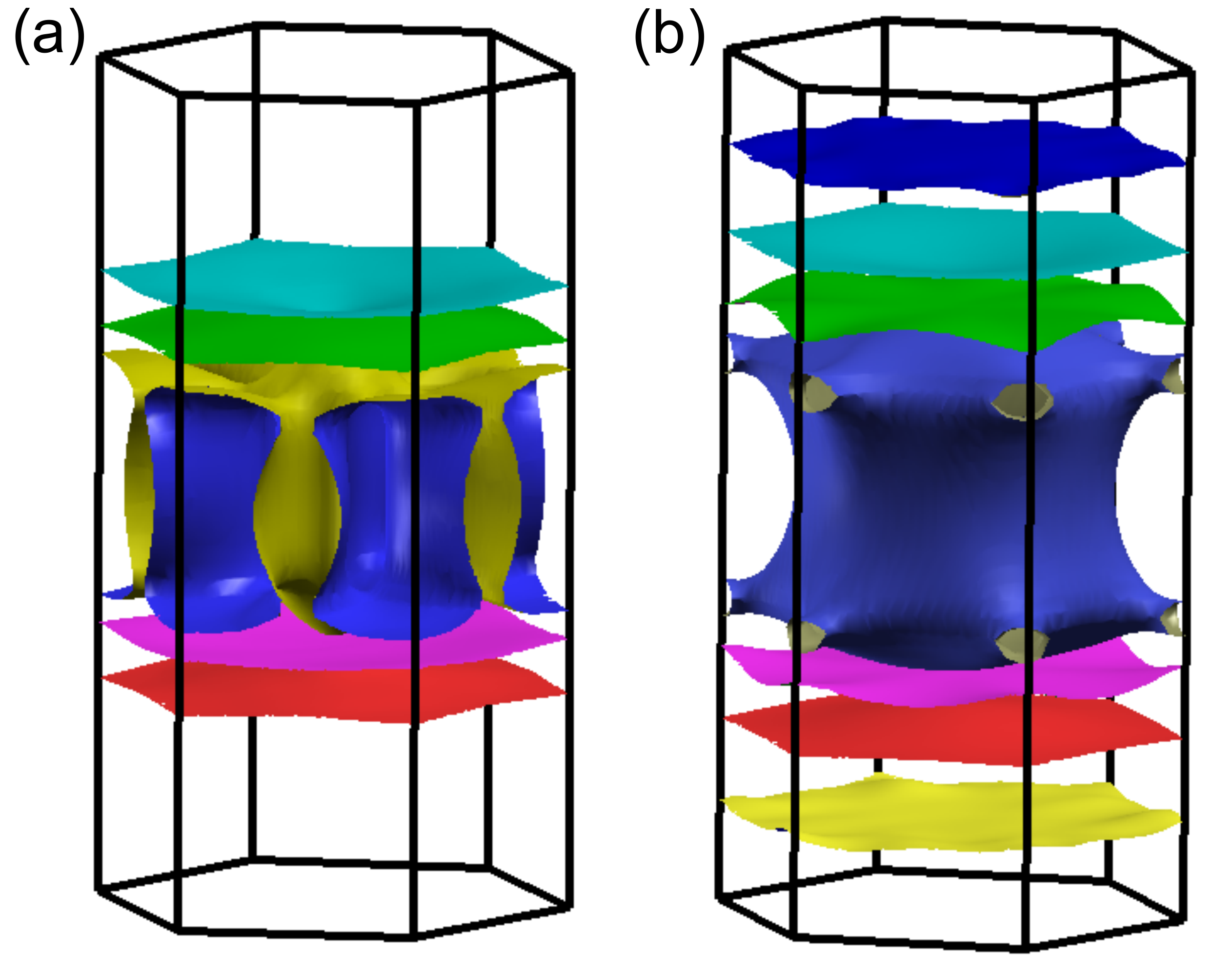}}
\caption{ Fermi surfaces of K$_2$Cr$_3$As$_3$ (a) and Rb$_2$Cr$_3$As$_3$(b) with experimental parameters in the paramagnetic state.
 \label{fermisurface} }
\end{figure}

The band structure and density of states(DOS) for K$_2$Cr$_3$As$_3$ with experimental structural parameters are shown in Fig.\ref{band} (a) and (b),(c), respectively. From the band structure and DOS, we find that Cr1 ions have more $d$ electrons than Cr2 ions, which is consistent with analysis from the crystal structure. It will lead to different magnetic moments at Cr1 and Cr2 sites in magnetic states. The Fermi surfaces are shown in Fig.\ref{fermisurface}(a), which are similar to those in Ref.\onlinecite{Jiang2014}. The band structure and DOS for Rb$_2$Cr$_3$As$_3$, shown in Fig.\ref{bandRb}, are similar to those of K$_2$Cr$_3$As$_3$. The main difference is that the 3D Fermi surface is quite different from that of K$_2$Cr$_3$As$_3$ and there is an additional electron Fermi surface near $A(0,0,\pi)$ point, as shown in Fig.\ref{fermisurface}(b).

\section{Spin excitations} \label{S3}

\begin{figure}[t]
\centerline{\includegraphics[height=7 cm]{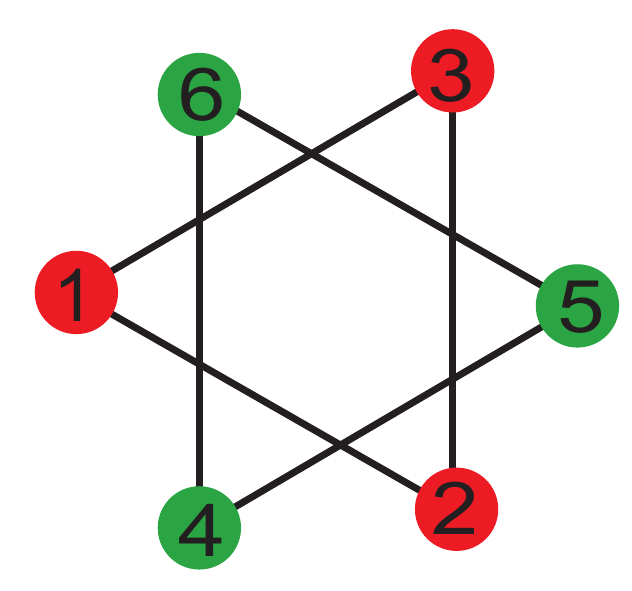}}
\caption{ Model of the Cr octahedral cluster in A$_2$Cr$_3$As$_3$.  \label{modelS} }
\end{figure}

\begin{figure}[t]
\centerline{\includegraphics[height=7 cm]{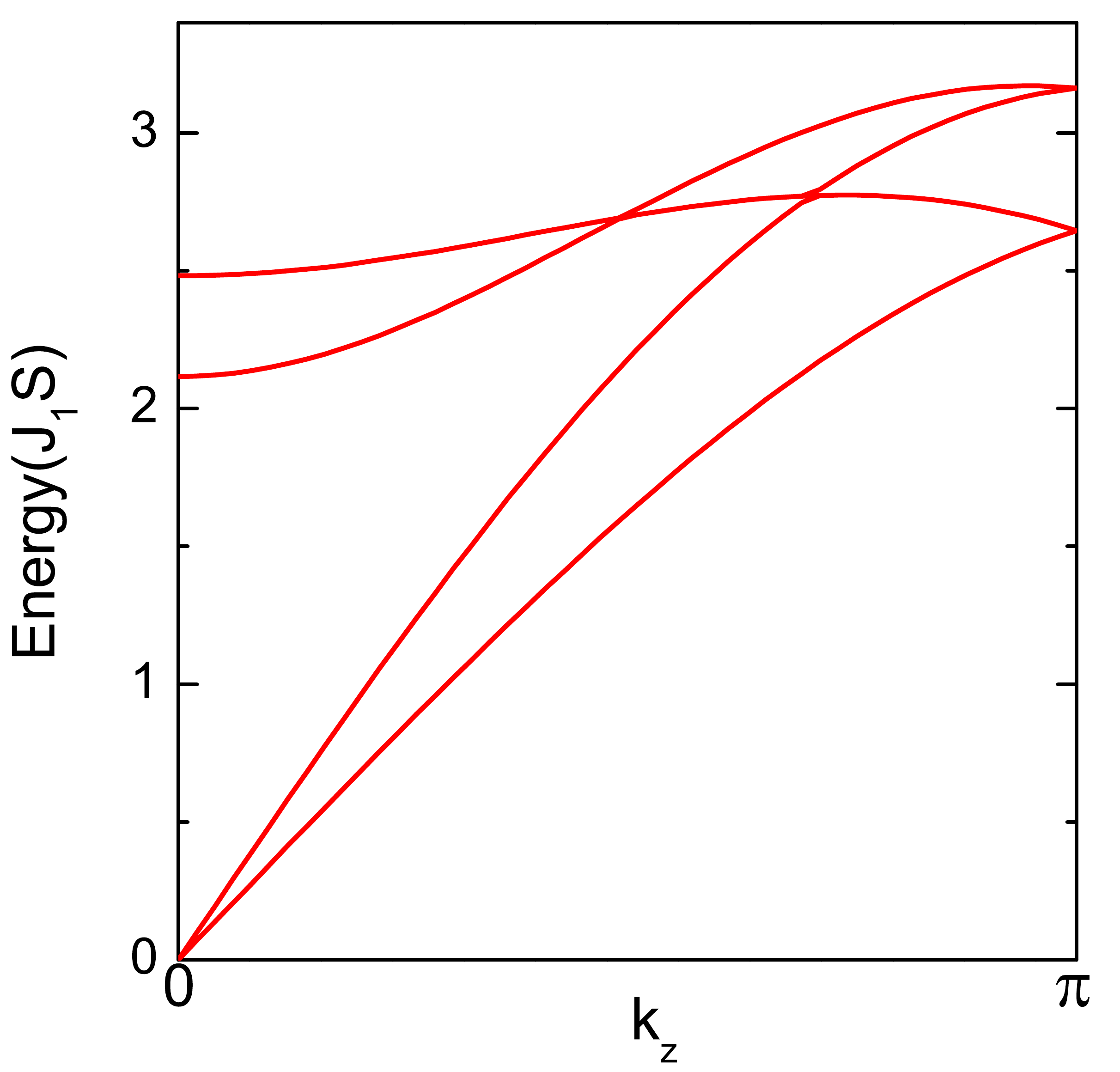}} \caption{ The calculated spin-wave dispersion of K$_2$Cr$_3$As$_3$ along $k_z$ direction, with $J_{1Cr1}=J_{1Cr2}=J_1$, $J'_1=0.7J_1$ and $J_2=-0.15J_1$.
 \label{spinwave} }
\end{figure}

The model of Cr octahedral clusters is shown in Fig.\ref{modelS}.  The magnetic excitations in the effective magnetic Hamiltonian, Eq.\ref{hm} , for A$_2$Cr$_3$As$_3$ can be calculated by using the Holstein-Primakoff transformation,
\begin{eqnarray}
 && S^{+}_{\alpha i}=\sqrt{2S-a^\dag_{\alpha i} a_{\alpha i}}a_{\alpha i},\nonumber \\
 && S^{-}_{\alpha i}=a^\dag_{\alpha i}\sqrt{2S-a^\dag_{\alpha i} a_{\alpha i}},\nonumber \\
 && S^z_{\alpha i}=S-a^\dag_{\alpha i}a_{\alpha i},
\end{eqnarray}
where $\alpha$ labels sublattice and $\alpha=(1,2,3,4,5,6)$ and $a_{\alpha i}$ is a bosonic operator. Then, keeping only the linear terms and using Fourier transformation, the magnetic Hamiltonian can be expressed as,
\begin{widetext}
\begin{eqnarray}
H_M&=&\sum_{k}H_0(1,2)+H_0(2,3)+H_0(3,1)+H_0(4,5)+H_0(5,6)+H_0(6,4)\nonumber\\
&+&\sum_{k}H_1(1,4)+H_1(1,6)+H_1(2,4)+H_1(2,5)+H_1(3,5)+H_1(3,6)\nonumber\\
&+&\sum_{k}H_2(1,1)+H_2(2,2)+H_2(3,3)+H_2(4,4)+H_2(5,5)+H_2(6,6),\\
H_0(1,2)&=&\frac{SJ_1}{4}(a^\dag_{1k} a_{2k}+ a_{1k} a^\dag_{2k})+\frac{SJ_1}{2}(a^\dag_{1k} a_{1k}+a^\dag_{2k} a_{2k})-\frac{3}{4}SJ_1(a^\dag_{1k} a^\dag_{2-k}+a_{1k} a_{2-k})-\frac{1}{2}NJ_1S^2,\\
H_1(1,4)&=&\frac{SJ'_1}{2}cos(\frac{k_z}{2})(a^\dag_{1k} a_{4k}+ a^\dag_{4k} a_{1k})-\frac{3}{2}SJ'_1cos(\frac{k_z}{2})(a_{1k} a_{4-k}+a^\dag_{1k} a^\dag_{4-k})\nonumber \\
&+&SJ'_1(a^\dag_{1k} a_{1k}+a^\dag_{4k} a_{4k})-J'_1NS^2,\\
H_2(1,1)&=&2SJ_2cos(k_z)(a^\dag_{1k} a_{1k})-2SJ_2cos(k_z)(a^\dag_{1k} a_{1k}+a^\dag_{1k} a_{1k})+J_2NS^2.
\end{eqnarray}

In the basis $\Phi^\dag(k)=(a^\dag_{1k},a^\dag_{2k},a^\dag_{3k},a^\dag_{4k},a^\dag_{5k},a^\dag_{6k},$ $a_{1-k},a_{2-k},a_{3-k},a_{4-k},a_{5-k},a_{6-k})$, the magnetic Hamiltonian is,
\begin{eqnarray}
H_M&=&\frac{1}{2}\sum_k\Phi^\dag(k)h(k)\Phi(k)-S(S+1)N(3J_1+6J'_1-6J_2),\\
h(k)&=&\left(\begin{array}{cccccccccccc}
A& B & B & D & 0 & D & 0 & C & C & E & 0 & E  \\
B& A & B & D & D & 0 & C & 0 & C & E & E & 0 \\
B& B & A & 0 & D & D & C & C & 0 & 0 & E & E \\
D& D & 0 & A & B & B & E & E & 0 & 0 & C & C \\
0& D & D & B & A & B & 0 & E & E & C & 0 & C \\
D& 0 & D & B & B & A & E & 0 & E & C & C & 0 \\
0& C & C & E & 0 & E & A & B & B & D & 0 & D \\
C& 0 & C & E & E & 0 & B & A & B & D & D & 0 \\
C& C & 0 & 0 & E & E & B & B & A & 0 & D & D \\
E& E & 0 & 0 & C & C & D & D & 0 & A & B & B \\
0& E & E & C & 0 & C & 0 & D & D & B & A & B \\
E& 0 & E & C & C & 0 & D & 0 & D & B & B & A \\
 \end{array}\right), \\
A&=&J_1S+2J'_1S-2J_2S+2J_2Scosk_z,\\
B&=&\frac{1}{4}J_1S,\\
C&=&-\frac{3}{4}J_1S,\\
D&=&\frac{1}{2}J'_1Scos(\frac{k_z}{2}),\\
E&=&-\frac{3}{2}J'_1Scos(\frac{k_x}{2}).
\end{eqnarray}
\end{widetext}
The spectrum of the spin wave is given in Fig.\ref{spinwave}, with $J'_1=0.7J_1,J_2=-0.15J_1$. In A$_2$Cr$_3$As$_3$, the bond lengths in Cr1As1 and Cr2As2 planes are different, which leads to different $J_1$ couplings in Cr1As1 and Cr2As2 planes. Therefore, we plot the spectrum of spin wave (Fig.\ref{phasespinwave}(b)) in the main text with $J_{1Cr1}=J_1,J_{1Cr2}=0.9J_1$ and $J'_1=0.7J_1,J_2=-0.15J_1$. Compared with Fig.\ref{spinwave}, there are gaps on the Brillouin Zone boundary, indicating the intrinsic two sublattices.

\end{bibunit}

\end{document}